\documentclass[10pt]{emulateapj}
\usepackage{apjfonts}
\usepackage{epsfig}
\usepackage{amsmath}
\usepackage{natbib}

\begin{document}

\title{Probing the environment of gravitational wave transient  sources with TeV afterglow emission}

\author{Qin-Yu Zhu, Xiang-Yu Wang\altaffilmark{1,2}}
\affil{$^1$ School of Astronomy and Space Science, Nanjing University, Nanjing 210093, China \\
$^2$ Key laboratory of Modern Astronomy and Astrophysics (Nanjing University), Ministry of Education, Nanjing 210093, China \\
}

\begin{abstract}
Recently, Advanced Laser Interferometer Gravitational-wave
Observatory (aLIGO) detected gravitational wave (GW) transients
from mergers of binary black holes (BHs). The system may also
produce a wide-angle, relativistic outflow if the claimed short
GRB detected by GBM is in real  association with GW 150914. It was
suggested that mergers of double neutron stars (or neutron
star-black hole binaries), another promising source of GW
transients, also produce fast, wide-angle outflows. In this paper,
we calculate the high-energy gamma-ray emission arising from the
blast waves driven by these wide-angle outflows. We find that TeV
emission arising from the inverse-Compton process in the
relativistic outflow resulted from mergers of binary BHs similar
to those in GW 150914 could be detectable by ground-based IACT
telescopes such as Cherenkov Telescope Array (CTA) if the sources
occur in { a dense medium with density  $n \ga 0.3\  \rm
cm^{-3}$}. For neutron star-neutron star (NS-NS) and NS-BH
mergers, TeV emission from the wide-angle, mildly-relativistic
outflow could be detected as well if they occur in a  dense medium
{with $n \ga 10-100\ \rm cm^{-3}$.} Thus TeV afterglow emission
would be a useful probe of the environment of the GW transients,
which could shed light on the evolution channels of the
progenitors of GW transients.

\end{abstract}


\section{Introduction}
The era of gravitational wave astronomy has begun with the first
science run of the recently upgraded LIGO  from 2015 September to
2016 January \citep{2016PhRvL.116f1102A}. The GW frequency range
that LIGO and Virgo are sensitive to is expected to be dominated
by mergers of compact stellar-mass objects that are most likely
remnants of stellar evolution: two neutron stars (NS-NS), two
black holes (BH-BH), or a NS and a BH. On 2015 September 14 at
09:50:45 UTC the LIGO collaboration detected the first GW
event--GW150914. The trigger was determined to be consistent with
a waveform predicted by General Relativity from the inspiral and
merger of two stellar-mass BHs, with  masses around $29 M_\odot$
and $36 M_\odot$, respectively \citep{2016PhRvL.116f1102A}.

\cite{2016AAS...22741602C} reported a tentative detection of a weak
transient gamma-ray source GW150914-GBM lasting  about 1 s, 0.4 s
after the LIGO trigger on GW150914. GW150914-GBM is consistent
with being due to a low-fluence short gamma-ray burst (GRB) at an
unfavorable viewing geometry to the GBM detectors, which is not
expected from a BH- BH merger. The isotropic-equivalent luminosity
of GW150914-GBM in the 1 keV to 10 MeV energy range was
$1.8^{+1.5}_{-1.0}\times 10^{49}{\rm erg s^{-1}}$ using a
$410^{+160}_{-180}$ Mpc distance inferred from the GW150914 event.
GW150914 was outside of the field of view of the Fermi Large Area
Telescope (LAT) initially and no GeV afterglow was detected when
it could observe \citep{2016ApJ...823L...2A}.

As the ground-based Imaging Atmospheric Cherenkov Telescopes
(IACTs) usually have a better sensitivity than Fermi/LAT, we here
study whether the TeV emission from the GW events could  be
detected by future IACTs, such as the planned Cherenkov Telescope
Array (CTA). It has been argued that CTA  is well suited to follow
up GW transients \citep{2014MNRAS.443..738B}. The short GRB possibly
associated with GW150914 should be produced by a wide-angle
outflow, since the chance of detection would be too low if it is a
strongly beamed jet. The outflow producing this short GRB must be
highly relativistic. It has been suggested that an accretion disk
may exist around  one of the BHs via the formation of the dead
zone \citep{2014ApJ...781..119P} in a  BH-BH merger system. This system
could then also have sub-relativistic or mildly relativistic wind
outflow if the disk produce a wind outflow \citep{2016ApJ...822L...9M}.

For the NS-NS mergers or NS-BH merges, significant wide-angle mass
outflows are also expected, as have been seen in the numerical
simulations. The mass of the outflow is in the range of
$M_{ej}=10^{-4}-10^{-2} M_\odot$ with a velocity of $v_w=0.1-0.3
c$. For NS-BH mergers, the ejecta mass can be up to $\sim 0.1
M_\odot$ \citep{2015PhRvD..92d4028K}.  A fraction of the mass may
have a higher speed or become relativistic with a Lorentz factor
$\Gamma_0 \sim 2$ and energy about $\sim 10^{49} {\rm erg}$ for
NS-NS mergers or $\sim 10^{50}\rm erg $ for NS-BH mergers (Nakar
\& Piran 2011). In this paper, we calculate the expected
high-energy gamma-ray emission produced by the blast waves driven
by these wide-angle outflows. The high-energy gamma-ray emission
from the small-angle, relativistic jets has been calculated in
literatures \citep{2014ApJ...787..168V,2014MNRAS.443..738B}. We
here study the TeV emission from the wide-angle,
mildly-relativistic outflows. Compared with the small-angle jets,
the wide-angle geometry can significantly increase the probability
for the follow-up electromagnetic detections of GW transients.

We present the calculation of the self inverse-Compton (IC)
emission for both relativistic  and mildly-relativistic outflows
in \S 2. In \S 3, we apply the result to both the BH-BH merger and
NS-NS (and NS-BH) merger scenarios. Finally, we give our
discussions and conclusions.

\section{The  IC emission from wide-angle outflows}

\subsection{The IC emission from the blast wave}
The short GRB possibly associated with GW150914, if real, should
be produced by a highly relativistic, wide-angle outflow. The
calculation for relativistic wide-angle outflow is quite similar
to the case of GRB jets except that the outflow may have a much
wider opening angle. We assume the relativistic outflow has an
initial Lorentz factor $\Gamma_0 \sim 100$ and it interacts with
homogenous ambient medium with a number density of $n$. Then the
blast wave driven by the outflow starts to decelerate at a time $
t_{dec}=(E/(32\pi n m_p \Gamma_0^8 c^5))^{1/3}=10 {\rm s}
E_{50}^{1/3}n^{-1/3} \Gamma_{0,2}^{-8/3} $ and becomes
non-relativistic at a time $ t_{nr}=(3 E/(4\pi n m_p
c^2))^{1/3}=65 E_{50}^{1/3}n^{-1/3} {\rm days} $, where $E$ is the
kinetic energy of the blast wave.

For the NS-NS/NS-BH mergers, numerical simulations show wide-angle
outflows  with mass $m=0.01-0.1 M_\odot$ and typical velocity $v =
0.1-0.5 c$.  Some works (e.g.,
\citealt{2006ApJ...648..510L,2008MNRAS.390..781M,2009ApJ...690.1681D}
) suggest the inner part of the outflow can be accelerated up to
relativistic velocities if the compact binaries form an accretion
disk before the coalescence. We assume that the
mildly-relativistic ejecta with $\Gamma \simeq 2$ have an energy
of about $10^{49}-10^{50}{\rm erg}$. Such amount of energy  is
consistent with the  non-detection of the late-time radio emission
in some compact binary mergers \citep{2016ApJ...819L..22H}. For
the mildly-relativistic outflow, the peak of the IC emission
occurs at the deceleration time, which is
$t_{dec}\simeq4.3\times10^5 {\rm s}\, E_{50}^{1/3}n^{-1/3}
(\Gamma_{0}/2)^{-8/3}$.

TeV emission is mainly produced by synchrotron self-Compton (SSC)
process of shock-accelerated electrons. The typical spectral
breaks in the SSC spectrum are \citep{2001ApJ...548..787S}
\begin{equation}
\nu_m^{IC}=2\gamma_m^2 \nu_m=3.36\times10^{14}{\rm Hz} f_p^4
\epsilon_{e,0}^4 \epsilon_{B,-2}^{1/2}E_{50}^{3/4}
n^{-1/4}t_6^{-9/4}
\end{equation}
and
\begin{equation}
\nu_c^{IC}=2\gamma_c^2\nu_c=6.03\times10^{26}{\rm Hz} \epsilon_{B,-2}^{-7/2}
t_6^{-1/4} E_{50}^{-5/4} n^{-9/4},
\end{equation}
where $p$ is the power-law index of the energy distribution of the
injected electrons,  $f_p=6(p-2)/(p-1)$, $\epsilon_e$ and
$\epsilon_B$ are respectively the equipartition factors of
magnetic field and electron energy, $\gamma_m=\epsilon_e
\frac{p-2}{p-1}\frac{m_p}{m_e}(\Gamma-1)$ and $\gamma_c$ are two
characteristic  Lorentz factors in the electron energy
distribution, and $\nu_m$ and $\nu_c$ are the corresponding break
frequencies \citep{1998ApJ...497L..17S}. The optical depth for IC
scatterings is $\tau=\frac{1}{3}\sigma_{\rm T} n R$ for a constant
density medium, where $R$ is the radius of the blast wave. Then
one can obtain the peak flux for the SSC component,
\begin{equation}
f_m^{IC}=\tau f_m= 1.08\times10^{-7}{\rm mJy} E_{50}^{5/4}n^{5/4}
\epsilon_{B,-2}^{1/2} t_6^{1/4} D_{L,27}^{-2},
\end{equation}
where $f_m$ is the peak flux of the synchrotron component and
$D_L$ is the luminosity distance of the source. The flux of the IC
component is given by
\begin{equation}
 f_{\nu}=\begin{cases}f_m^{IC} (\frac{\nu}{\nu_m^{IC}})^{-\frac{p-1}{2}} \ \ \ \ \ \ \ \ \  \mbox{, $\nu_m^{IC}< \nu<\nu_c^{IC}$}\\
f_m^{IC} (\frac{\nu_c^{IC}}{\nu_m^{IC}})^{-\frac{p-1}{2}}
(\frac{\nu}{\nu_c^{IC}})^{-\frac{p}{2}}\ \ \ \ \mbox{,
$\nu_c^{IC}< \nu$}.
\end{cases}
\end{equation}
At high energies, the SSC spectrum could be affected by the
Klein-Nishina effect \citep{2009ApJ...703..675N,
2010ApJ...712.1232W}, which occurs at frequencies above a critic
frequency of
\begin{equation}
\nu_c^{KN}=\gamma_c m_e c^2=2.4\times10^{25}{\rm Hz}
\epsilon_{B,-2}^{-1} E_{50}^{-3/8} n^{-5/8}t_6^{1/8}.
\end{equation}
Then the flux can be obtained by using Eq.50 in \cite{2009ApJ...703..675N}.

To illustrate the TeV afterglow behavior, we show an example of
the light curves  at $h\nu={1 \rm TeV}$ by taking some reference
values for the parameters of the wide-angle outflows, which is
shown in Fig.1. The relativistic wide-angle outflow generates a
decaying TeV afterglow. Meanwhile, the mildly-relativistic outflow
and sub-relativistic outflow produce a rising light curve before
the deceleration time and a decay one after that. {Before
deceleration, since the blast wave velocity is constant, the
typical energy of electrons ($\gamma_m$) and the peak frequency
($\nu_m^{IC}$) are constant. As the number of radiating electrons
is increasing with time,  the inverse-Compton TeV flux increases
with time. After deceleration, as the blast wave velocity
decreases, the power of the inverse-Compton emission per electron
decreases rapidly with time, so the TeV flux starts to decrease
with time.}

Our calculations are based on the assumption of the electron
distribution of $p = 2.2$.  A larger $p$ may suppress the high
energy gamma-ray emission. This is because the TeV band is much
larger than the peak frequency of SSC emission $v_m^{IC}$ and even
larger than the cooling frequency $v_c^{IC}$ in some cases, which
makes the ratio between the TeV flux and the peak SSC flux
affected by the electron distribution.

\subsection{Detectability by CTA }
We first need to estimate the sensitivity of CTA.  According to
the performance of
CTA\footnote{https://portal.cta-observatory.org/Pages/CTA-Performance.aspx},
CTA South has a  differential sensitivity  of
$E^2\phi_0=1\times10^{-13}\rm erg\ cm^{-2} s^{-1}$ at energy of $1
\rm TeV$  for the $T_0=50\rm h$ exposure. Following the method in
\cite{2007ApJ...668..392G}, the sensitivity for an exposure time
$t_{exp}$ can be determined by the minimum fluence $F(t_{exp})$
with which a source is detecable \citep{2014MNRAS.443..738B}, i.e.
$F(t_{exp})\approx \kappa \phi_0 (\frac{T_0}{t_{exp}})^{1/2}
E^2_{cut}t_{exp} $,  where $E_{cut}$ is the cutoff energy of the
energy spectrum, and the constant $\kappa\leq 1 $ depends on the
width of the energy bin, the flux from the source and the spectral
shape of the detector sensitivity. Hereafter, we conservatively
assume $\kappa = 1$.

To determine the direction of a GW event, at least two separate GW
detectors are needed,  taking advantage of the different arrival
time of the GW signals at different locations
(\citealt{2016LRR....19....1A} and references therein). Based on
the results of \cite{2011CQGra..28j5021F},
\cite{2014MNRAS.443..738B} estimated the typical localization sky
area for LIGO Hanford and LIGO Livingstone network to be $\Omega
\approx 2000\ \rm
deg^2(\frac{8}{SNR})(\frac{erf^{-1}(CL)}{erf^{-1}(0.9)})$ {where
${erf}^{-1}(CL)$ is the inverse error function at certain
confidence level (CL) and we normally choose a confidence level of
$0.9$ ($90\%$) .} For the three detector case, the localization
sky area could be reduced to $\Omega \approx 100\ \rm
deg^2(\frac{8}{SNR})(\frac{erf^{-1}(CL)}{erf^{-1}(0.9)})$ or even
better($< 20 \rm deg^2$) at some special locations. For the medium
energy range of CTA, from $100\rm GeV$ to $10\rm TeV$, CTA has 40
telescopes with a Field-of-View of about $38\ \rm deg^2$.
Therefore, CTA can cover the required area determined by three (or
more) GW detectors with only 1-3 follow-up observation(s).
According to  \cite{2014MNRAS.443..738B}, we find that the
sensitivity in the survey mode is $S_{survey}\geq 0.5\ S_{det}$,
where $S_{det}=F(t_{exp})/t_{exp}$ is the single-pointing
detection threshold of CTA. Hereafter, we use $S_{det}$ as the TeV
sensitivity of CTA approximately.

As CTA can slew to a certain position within 100 seconds, it can
observe the early stage of the TeV afterglow if the GW transient
is localized, for example, by a short GRB. For the
mildly-relativistic outflow case, even if the spatial error is as
large as $100\ \rm deg^2$, CTA can cover the error in the survey
model with a few consecutive observations.  The sensitivity light
curve of CTA in the point observation mode is shown in Fig.1.  It
is worth noting that the CTA limit observation time is $t_{ob}\leq
50 \rm h$ because of observing strategies, so we fix the
sensitivity when $t>50 \rm h$.

\section{TeV afterglows in specific models }
\subsection{BH-BH mergers}
The possible  short GRB associated with GW150914 suggests that
BH-BH mergers may produce wide-angle, relativistic outflows with
an isotropic energy of $E\sim10^{50}{\rm erg}$
\citep{2016AAS...22741602C}. We study the conditions under which
the TeV afterglow from this relativistic outflow can be detected
by CTA. Fig.2 shows the TeV flux as a function of the density $n$
and the magnetic field equipartition factor $\epsilon_B$, as these
two parameters are the mostly unknown parameters. The distance of
the source is taken as $D_L=10^{27}{\rm cm}$, comparable to the
distance of GW150914 \citep{2016PhRvL.116f1102A}. Since the TeV
light curves show a decaying feature (Fig.1), the time at which
CTA starts observation, $t_{start}$, is important for TeV
detection. A conservative observational delay of $t_{start}\geq100
\rm s$ can be obtained because of a 1-minute delay from data
analysis/detection and ~1/2 minute delay from the maneuver of CTA
\citep{2014MNRAS.443..738B}. We show how the value of $t_{start}$
affects  CTA detection  by choosing two values of $10^3{\rm s}$
and $10^4{\rm s}$ respectively, as shown respectively by the left
and right panel of Fig.2. We find that the TeV afterglow emission
for the case of $t_{start}=10^4 {\rm s }$ can be detected by CTA
when $n\ga 1{\rm cm^{-3}}$ for $\epsilon_B\ga 10^{-3}$, while for
the case of $t_{start}=10^3 {\rm s }$,  $n\ga 0.3\ {\rm cm^{-3}}$ is
needed.

\subsection{NS-NS / NS-BH mergers}
The mergers of NS-NS or NS-BH binaries are promising sources for
gravitational detection by aLIGO and aVirgo. Short GRBs, resulted
from highly relativistic jets, are believed to be produced by such
systems. As the opening angle of the relativistic jets are usually
small, perhaps of $\theta_j\sim 0.1$ \citep{2014ARA&A..52...43B},
the jet direction only covers a very small fraction of the sky.
Here we pay attention to the wide-angle outflow from NS-NS (NS-BH)
mergers. Numerical simulations of compact binary merger  systems
have been carried out by many groups (e.g.
\citealt{1999A&A...341..499R,2000A&A...360..171R,2005ApJ...634.1202R,2001A&A...380..544R,2008PhRvD..78f4054Y,2010CQGra..27k4105R,2010PhRvL.104n1101K}).
The mergers give rise to unbound matter ejection through dynamic
processes. The mass ejected depends primarily on the total binary
mass, the mass ratio and the equation of state. In some
simulations, disk formation can contribute to several outflow
sources. For example, neutrino heating drives a wind from the disk
surface (e.g.,
\citealt{2006ApJ...648..510L,2008MNRAS.390..781M,2009ApJ...690.1681D}).
The outflow velocity may become relativistic for winds that are
ejected from close to the central object. Thus, similar to
\cite{2011Natur.478...82N}, we assume that a fraction of the mass
becomes relativistic with a Lorentz factor of $\Gamma \sim 2$ and
energy about $\sim 10^{49} {\rm erg}$ for NS-NS mergers or $\sim
10^{50}\rm erg $ for NS-BH mergers. If the central object after
the merger of double NSs is a fast-rotating Magnetar, the energy
in the mildly-relativistic ejecta would be even larger
{\footnote{Wang et al. (2016) have studied the SSC emission from
the reverse shock powered by post-merger millisecond magnetars.}}
\citep{2004ApJ...611..380T}.

We calculate the TeV afterglow emission at the deceleration time
of the mildly-relativistic outflow, which is shown in Fig.3. In
this case, the light curve reaches a peak at the time $t_{dec}\sim
3.0\times10^{5}E_{49}^{1/3}n^{-1/3} \rm s$. The duration of the
peak lasts several days, so CTA can scan the whole error box of
the GW transient in this period. According to Fig.3, the detection
of a NS-BH merger at $D_L=10^{26.5}{\rm cm}$ by CTA requires that
the density should be $n\ga 10 {\rm cm^{-3}}$ for a wide range of
$\epsilon_B$ ($10^{-4}\la \epsilon_B\la 10^{-2}$). Such a high
density is possible if the merger occurs in the disk of starburst
galaxies. We note that some short GRBs have been found to occur in
such star-forming galaxies \citep{2014ARA&A..52...43B}. Detection
of TeV afterglow would strongly support that mergers  indeed occur
in the star-forming regions.

\section{Conclusions and discussions}

We considered the TeV afterglow emission from the wide-angle,
relativistic outflow of BH-BH mergers and from the
mildly-relativistic outflow of NS-NS/NS-BH mergers. We found that
TeV afterglow emission could be detected if the compact binary
mergers occur in a dense ambient environment({i.e., $n\ga 0.3\
{\rm cm^{-3}}$ for GW150914-like binary BH mergers and $n\ga
10-100 {\rm cm^{-3}}$ for NS-NS/NS-BH mergers}). For the binary
black-hole merger GW150914, it has been suggested that the binary
black-holes formation  involves either isolated binaries in
galactic fields or dynamical interactions in young and old dense
stellar environments \citep{2016ApJ...818L..22A}.  The  binary
black-hole progenitor of GW150914  could  have formed in the local
universe with a short merger delay time, or it could have formed
at a higher redshift with a long merger delay time. In the
short-delay merger scenario, the mergers are expected to  occur in
the star-forming regions, so the ambient medium density is high.
Detections of TeV afterglow emission by CTA in future would
support the short-delay merger scenario for GW events like
GW150914.

For the evolution of NS-NS or NS-BH binaries, short GRB
observations have shed some light on the burst environment.  Short
GRBs occur in both elliptical and star-forming galaxies, with the
latter accounting for about $80\%$ of the sample (Berger 2014).
For mergers occurring in  dense star-forming regions with a high
gas density, TeV afterglow emission could be detected by CTA. If a
Magnetar forms after the merger of double NSs, the chance of
detection of TeV afterglow would be increased as the energy in the
mildly-relativistic outflow is increased by the injection energy
from the central Magnetar.

The  mildly-relativistic ejecta resulted from NS-NS or NS-BH
mergers will also produce a long term radio emission due to its
interaction with the surrounding ISM. Recently,
\cite{2016ApJ...819L..22H} search for such emission from two short
GRBs, i.e. GRB 130603B and GRB 060614, with radio telescopes, but
find non-detections. The upper limit flux of the radio emission
can nevertheless  put useful constraints on the parameters of the
ejecta and density of the surrounding medium. To study the
constraints, we calculate the radio emissions at { 6.7 GHz} from
the mildly-relativistic ejecta with different parameters adopted,
which are shown in Fig.4. We find that the radio emission violates
the upper limits only when both the ISM density is higher than
$\sim100 {\rm cm^{-3}}$ and the ejecta energy is higher than
$10^{50}{\rm erg}$. Therefore, a wide parameter space still exists
for which TeV afterglow could be detected by CTA while meanwhile
the radio constraints are satisfied.

\acknowledgments We thank Liang-Duan Liu and Yuan-Pei Yang for
helpful discussions. This work is supported by the National Basic
Research Program (973 Program) of China under Grant No.
2014CB845800,   the National Natural Science Foundation of China
under  Grants  No. 11273016, and the Basic Research Program of
Jiangsu Province under Grant No. BK2012011.

\clearpage

\begin{center}

\begin{figure}
\includegraphics[scale=0.4]{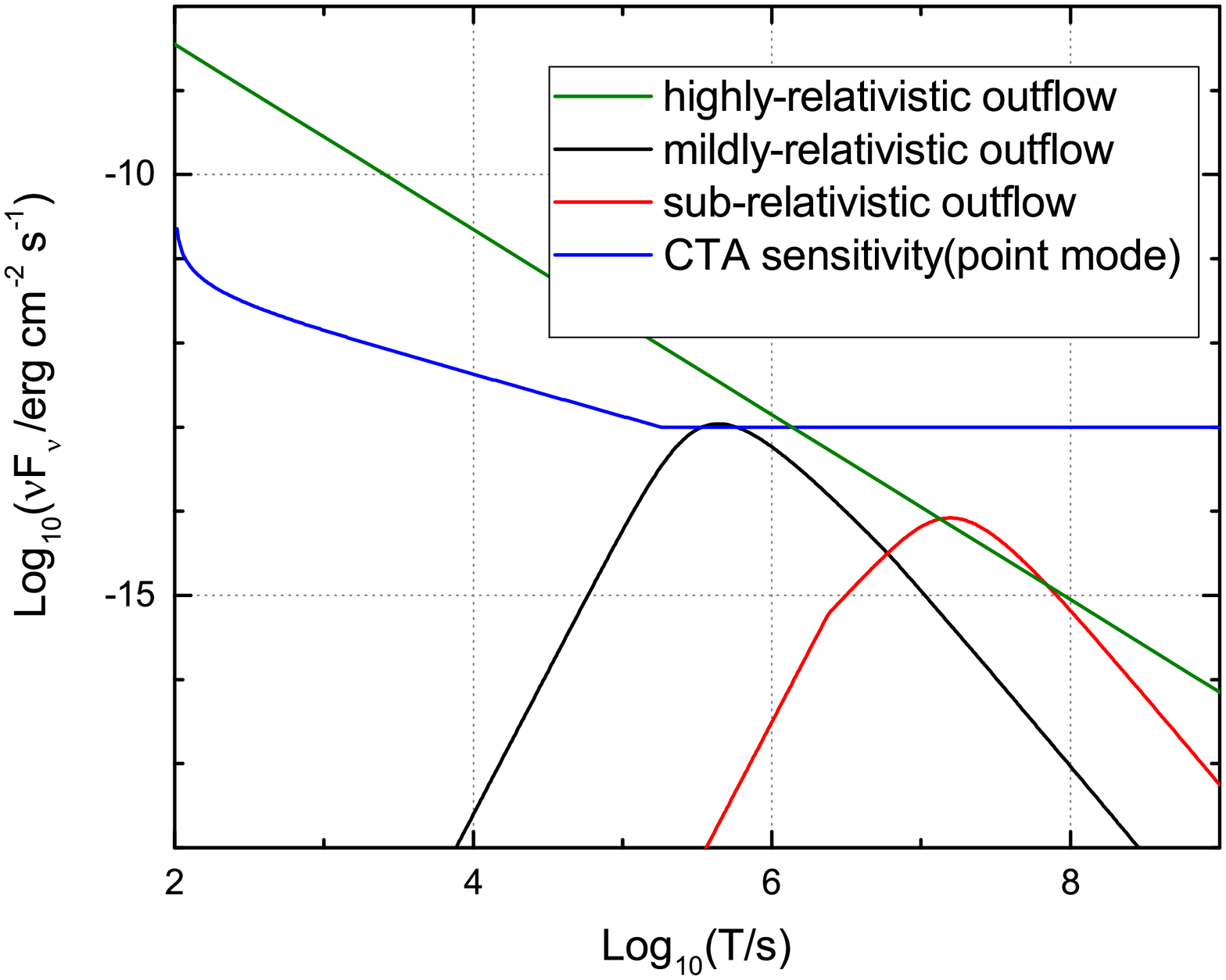}%
\hfill \caption{TeV light curves in the relativistic outflow
scenario (the green line), mildly-relativistic outflow scenario
(the black line) and sub-relativistic one (the red line). The
green line is calculated with the parameters $E=10^{50}{\rm
erg},\Gamma=100$. The black line is calculated with the parameters
$E=10^{50}{\rm erg},\Gamma=2$. The red line is calculated with the
parameters $E=3\times10^{52}{\rm erg},v=0.3c$. Other parameters
used in the calculation are fixed to
$\epsilon_e/0.4=\epsilon_{B,-2}=n=D_{L,26.5}=p/2.2=1$. The blue
line is the CTA sensitivity in the pointing observing mode.}
\end{figure}

\begin{figure}
\includegraphics[scale=0.3]{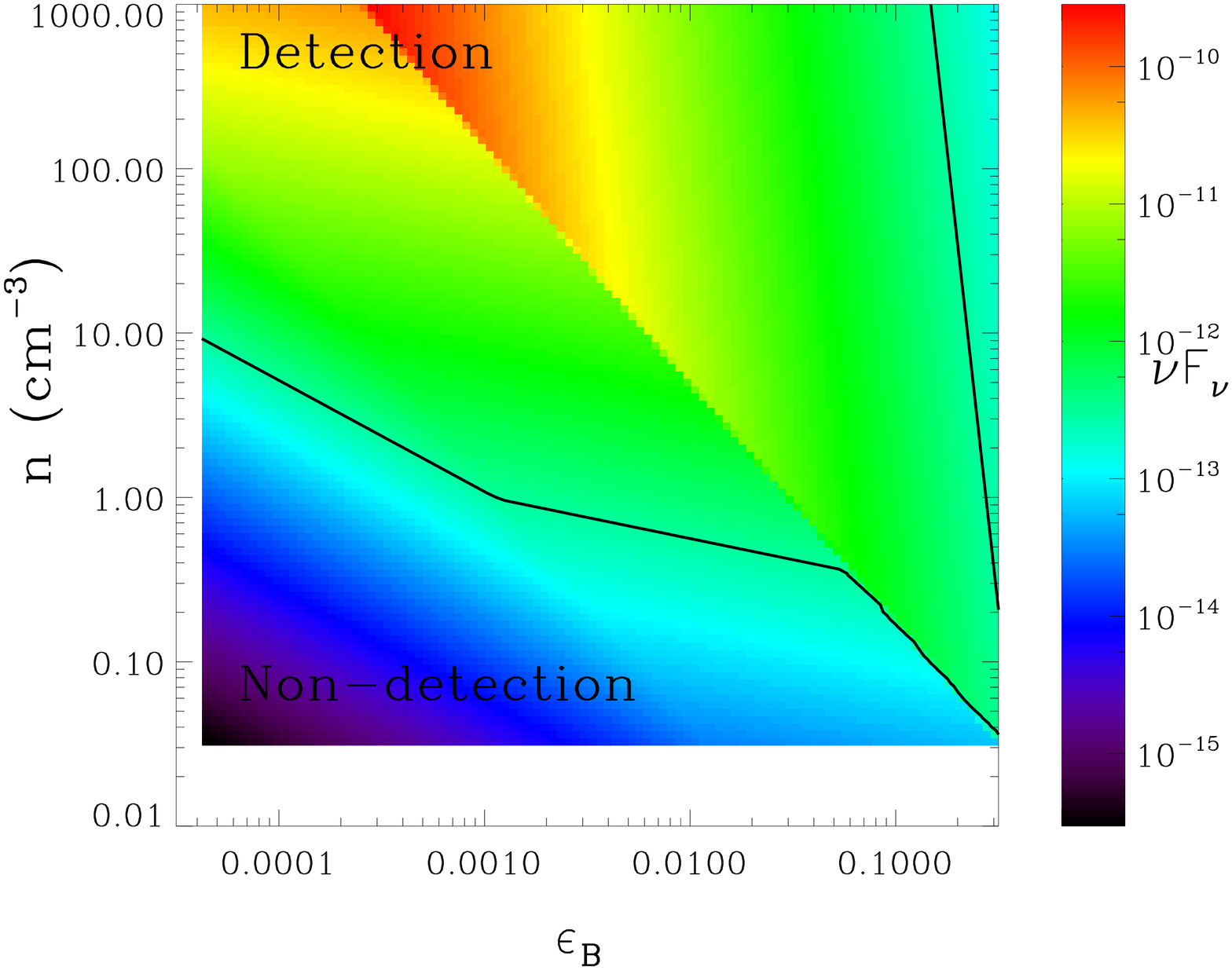}%
\includegraphics[scale=0.3]{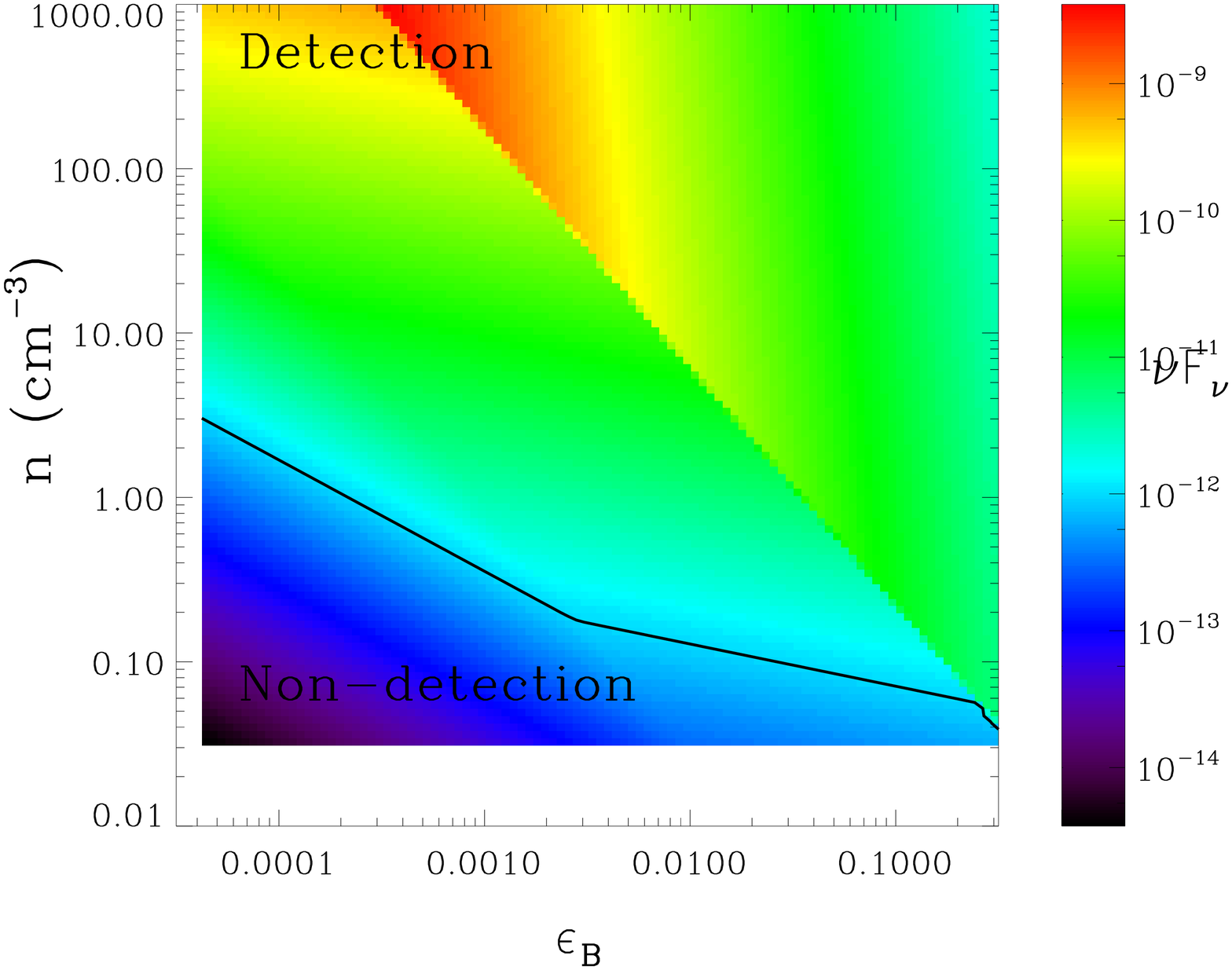}%
\hfill \caption{TeV flux in the relativistic outflow scenario. The
fluxes  are represented by colors in the unit of $\rm erg\ cm^{-2}
s^{-1}$. We fix $p=2.2,\Gamma=100,  \epsilon_e=0.4, E = 10^{50}
\rm erg$ and  $D_L=10^{27}\rm cm$.  The left panel is calculated
for $t_{start}=10^4\rm s$ and the black line represents CTA
sensitivity of 5-hour observation in the point observation mode.
The right panel shows the case with $t_{start}=10^3 \rm s $ and
for CTA 0.5-hour observation sensitivity.}
\end{figure}

\begin{figure}
\includegraphics[scale=0.3]{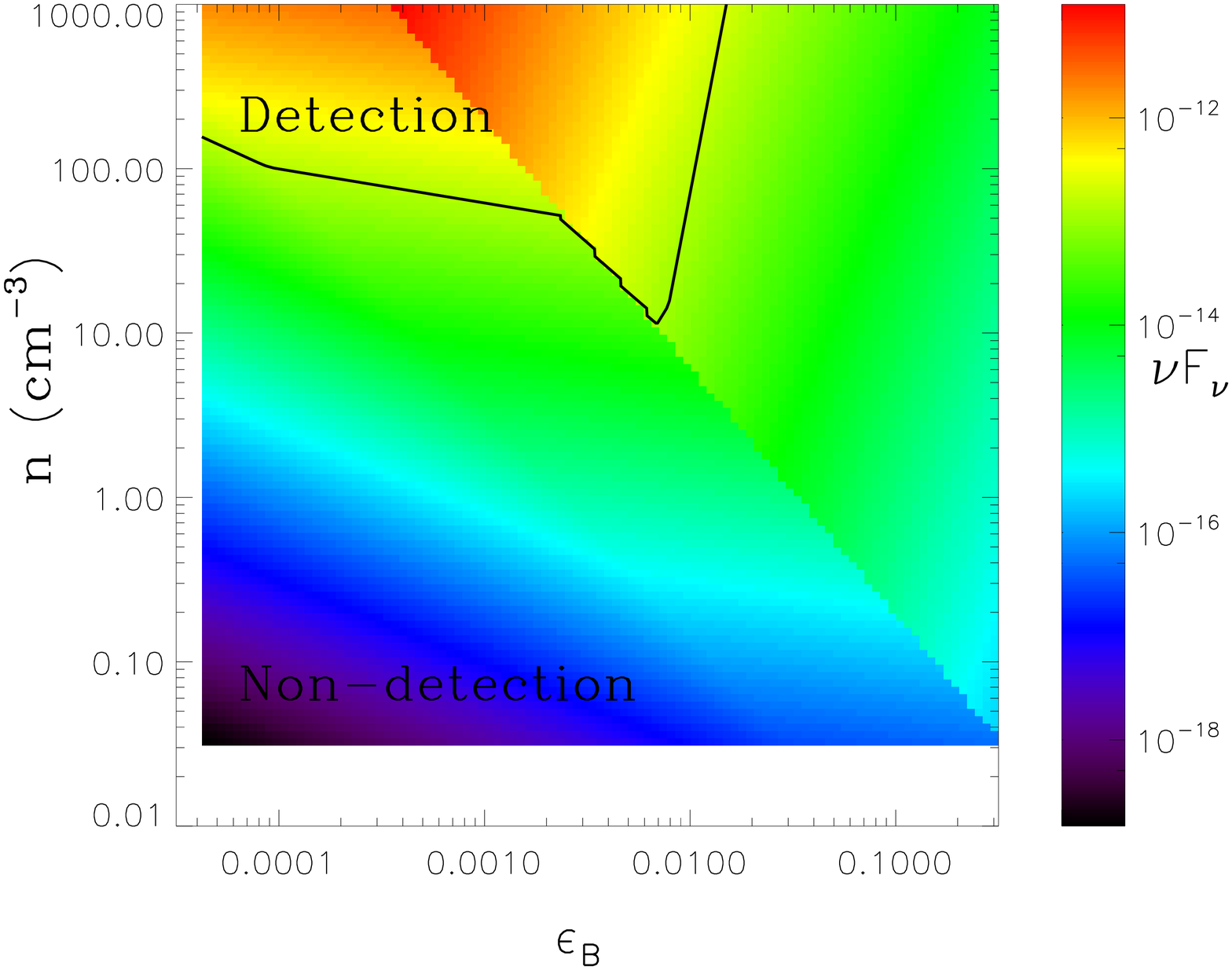}%
\includegraphics[scale=0.3]{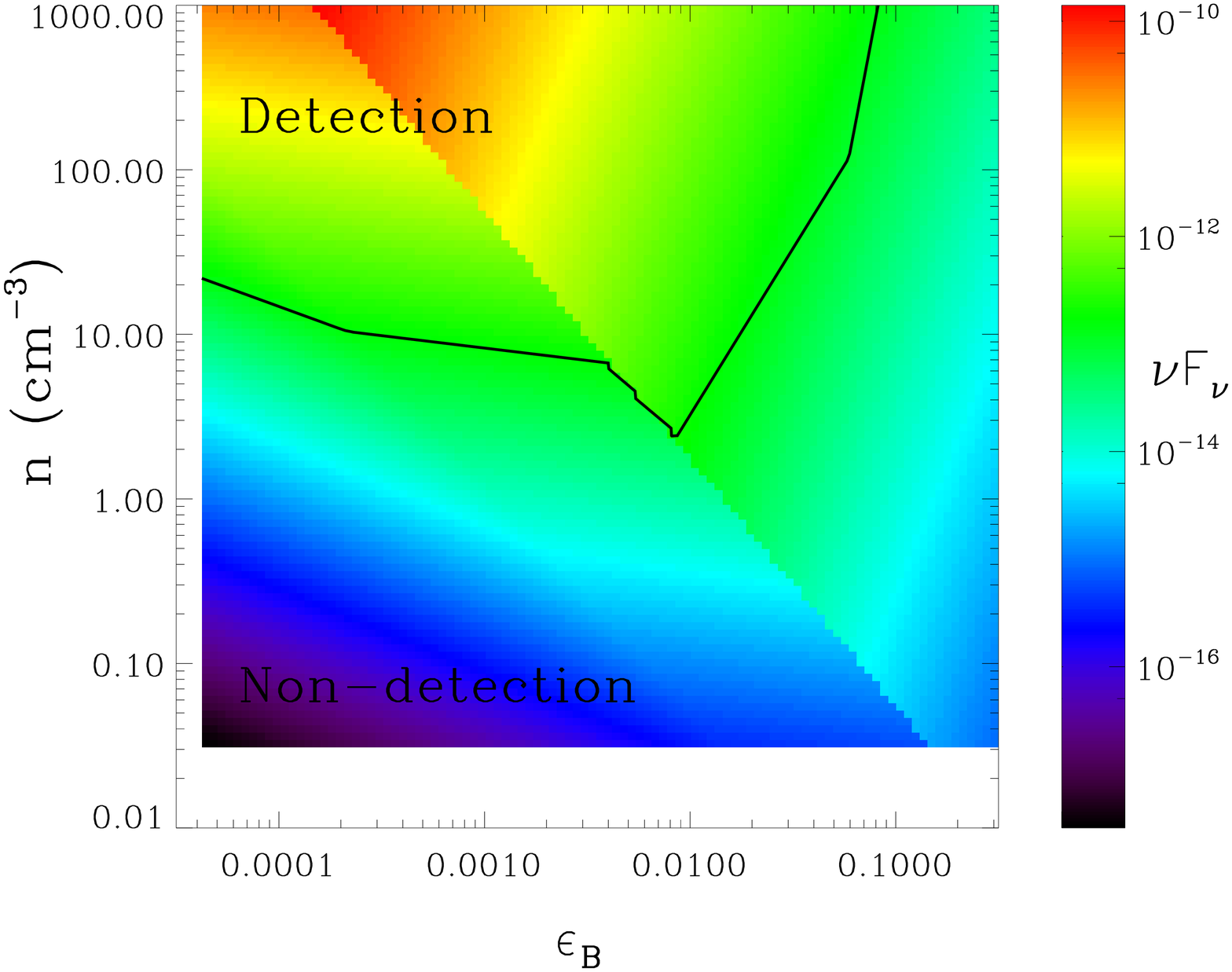}\\%
\caption{TeV flux in the mildly-relativistic outflow scenario for
NS-NS mergers (left panel) and NS-BH mergers (right panel). The
kinetic energis of the blast waves are assumed to be
$E=10^{49}{\rm erg}$ and  $E=10^{50}{\rm erg}$ for NS-NS mergers
and NS-BH mergers respectively.  Other parameters uased are
$p=2.2, \epsilon_e=0.4, \Gamma=2,  D_L=10^{26.5}\rm cm$. TeV flux
are represented by colors in the unit of $\rm erg\ cm^{-2}
s^{-1}$. The black line shows the sensitivity of CTA 50-hour
single-pointing observation.}
\end{figure}

\begin{figure}
\includegraphics[scale=0.4]{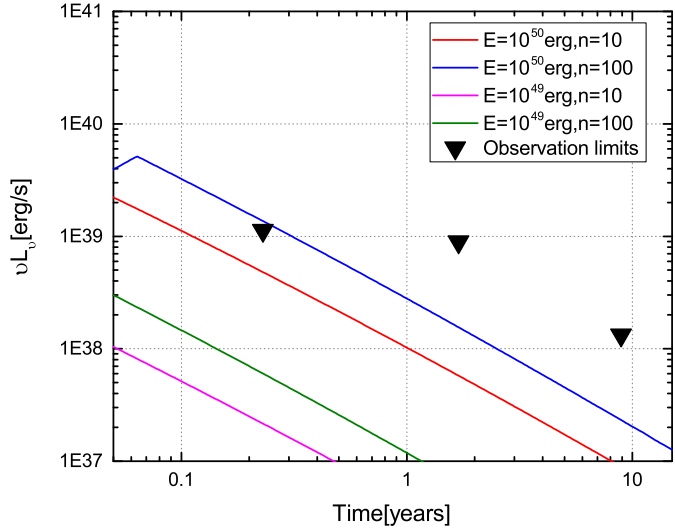}%
\hfill \caption{Radio emissions at {6.7 GHz} from the
mildly-relativistic outflow in comparison with the upper limits
for GRB 130603B and GRB 060614. Different lines represent
different values of number density $n$ and ejecta energy $E$,
while other parameters are fixed at $p=2.2$, $\epsilon_e=0.4$,
$\epsilon_B=0.01$, $\Gamma=2$ and  $z=0.356$ (the redshift of GRB
130603B). Solid triangles represent the radio upper limits flux
taken from Horesh et al. (2016). }
\end{figure}

\end{center}
\end{document}